# Quickpie: An interface for Fast and Accurate Eye Gaze based Text Entry

**Pawan Patidar**
School of Information Technology
IIT Kharagpur
ppatidar77@gmail.com

**Himanshu Raghuvanshi**
Department of Information Technology
NIT Durgapur
himanshu.2334@gmail.com

**Sayan Sarcar**
School of Information Technology
IIT Kharagpur
mailtosayan@gmail.com

**ABSTRACT**
Pie menus are suggested as powerful tool for eye gaze based text entry among various interfaces developed so far. If pie menus are used with multiple depth layers then multiple saccades are required per selection of item, which is inefficient because it consumes more time. Also dwell time selection method is limited in performance because higher dwell time suffers from inefficiency while lower one from inaccuracy. To overcome problems with multiple depth layers and dwell time, we designed Quickpie, an interface for eye gaze based text entry with only one depth layer of pie menu and selection border as selection method instead of dwell time. We investigated various parameters like number of slices in pie menu, width characters and safe region, enlarged angle of slice and selection methods to achieve better performance. Our experiment results indicates that six number of slices with width of characters area 120 px performs better as compared to other designs.

**INTRODUCTION**
Eye gaze based interaction enables user to focus at point of interest on the display by just looking at it. To perform such cursor movements with eye gaze we need a camera which captures images of eye and an eye tracking system is required which takes these images as input and gives coordinates of point where user is looking as output. This way of interaction is particularly very useful for motor handicapped patients because eye movement is fastest motor movement and a person can perform with rotation speed of 700°/s [1].

To enter a letter user have to first focus on letter on virtual interface and then take some distinguished action. Two most obvious actions are blinking or dwelling. People blink involuntarily in every several seconds [2], so that it becomes difficult to distinguish between a desired and unintentional blink. One possible way to overcome this problem is to use prolonged blink. In dwelling mechanism, user has to fixate eye gaze on desired area for a specific amount of time called dwell time. Dwell time commonly varies from 200 to 1000 ms and it must be longer than the normal viewing time to prevent false selections. But longer dwell time results in significant degradation in performance, i.e. typing speed because it elapses for each selection of characters. Hence dwell time should be moderate to improve typing speed and avoid false selections according to needs and expertize of the user [6]. Therefore it is observed that blinking and dwelling both suffered from Midas touch problem.

Between two consecutive fixations our gaze suddenly jumps from one point to another through ballistic eye movements. Such movements between fixations are called saccades. A saccade commonly takes 30 to 120 ms having an amplitude range between 1° and 40°(average 15° to 20°) [1, 4]. Latency period of at least 100 to 200 ms occurs before eye moves to next area of interest and after a saccade eyes will fixate on an object between 200 and 600 ms [1, 4]. Therefore a selection of character can be done faster if it requires only one saccade or more saccades in same direction. Such sequence of saccades is very helpful if user is expert and can predetermine path to select a desired character. User can "mark ahead" path corresponding to a character without searching for it [5]. Therefore searching time can be significantly minimized.

Even when a user wants to fixate on an object by looking steadily, the eyes make small jittery motions. These jittery movements are of less than one degree and can be high frequency tremor or low frequency drift. Moreover high-acuity region of our retina called fovea covers approximately one degree of visual arc that is why we cannot determine precisely where user is pointing by looking steadily.

Considering above difficulties we designed the interface to fulfill following objectives:

1. Removing dwell time and minimizing searching time and eye movements to maximize typing speed.
2. Preventing false selections (Midas touch problem) to minimize error rate.

**DESIGN OF INTERFACE**
Quickpie consists of four regions: pie, characters region, safe region and selection region (see figure 1). A typical design has following properties:

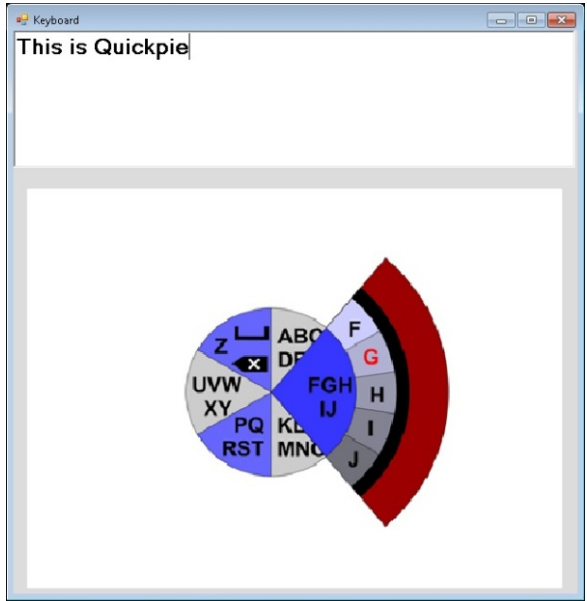

**Figure 1: Quickpie interface**

**Pie region**
The pie menu at first layer is having radius of 240 px and divided into six slices. Each slice is having initial angle (α) equal to 60° colored gray and blue alternatively. Each slice having five characters arranged in alphabetical order (except last slice which contains SPACE and CLEAR with z). When user focus on a slice then it is highlighted and its angle expands 20° in both clockwise and anticlockwise direction so that its enlarged angle (β) becomes 100°. β must be much less than 3×α otherwise major area of adjacent slices would be overlapped it would be difficult to focus on them.

**Characters region**
This is colored gray (darkness is increasing for each character in clockwise direction) and placed outside the border of pie. This region is having width of 100 px. Characters are arranged alphabetically in clockwise direction and each character is having angle(ϒ) equal to enlarged angle/number of characters per slice = β/5 = 20°. When user points on a particular character then its text color changes to red.

**Safe region**
This is black colored region at outer edge of characters region. This region is having width of 20 px and nothing happens when user focus on this region. This is used to minimize errors due to jittery movements.

**Selection region**
This is colored red and placed at outer edge of safe region. Width of this region doesn't is not important and its angle is equal to enlarged angle (β). When user enters in this region from safe region than its color turns to green and highlighted character in characters area is entered.

**OPERATION**
To enter a character, user has to follow following procedure:

1. Focus on the slice containing that character (Corresponding characters, safe and selection areas will appear immediately without any delay. If another slice was focused previously then its corresponding areas will disappear).

2. Focus on area of desired character (Text color of that character will be turned to red).

3. Cross safe region and enter into selection region(after crossing to selection region its color will turn to green, character selected in step 2 will be entered and displayed at current position of cursor so that there is no need to verify entered character by looking at the output text area).

If next character to be entered is in same slice then user do not need first step because desired character is already visible. Hence only second and third steps are required until the next character to be entered is in different slice (see Figure 2).

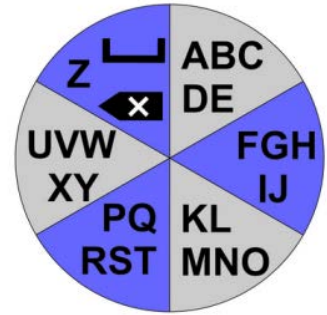

a. **Initial state of interface**

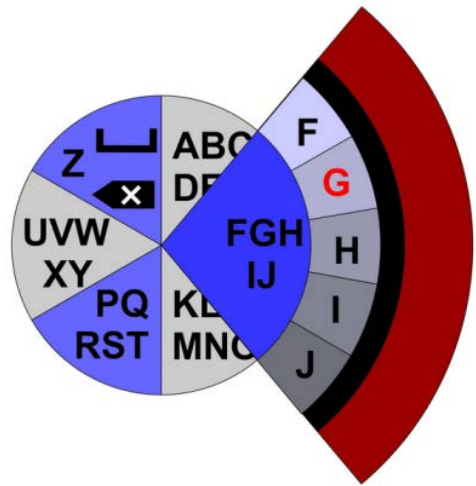

b. **Pointing G after pointing slice FGHIJ(G is selected)**

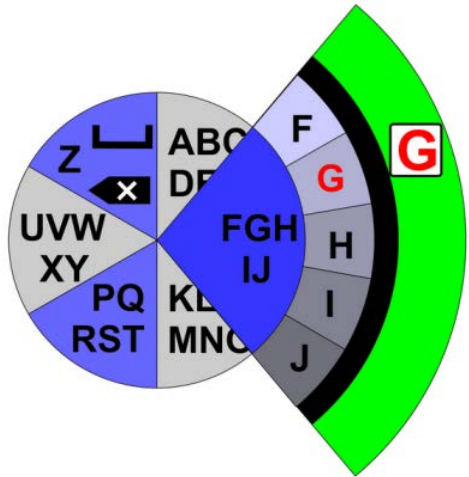

c. Crossed through safe region to selection region (G is entered)

**Figure 2: Sequence of steps to type G.**

Temporal threshold is totally removed because no dwell time is required in any of above three steps which is main advantage of this interface. Since $\Upsilon = \beta/5$, $\beta$ should be taken much higher than $\alpha$ because large items are easy and accurate to focus [7]. Increasing radius of pie is another way to enlarge size of each character but eye movements would be higher in that case which may result in performance degradation. Novice users may takes longer searching time but expert user can search fast and can predetermine path to be followed to enter a sequence of characters. This process of prediction is called "mark ahead" [5], which is another main advantage of this interface.

Characters area, safe area and selection area are in same direction so that a character can be entered in only one saccade or more saccades in nearly same direction. Therefore it eliminates disadvantage of delays which occurs between saccades in different directions. Moreover characters are alphabetically arranged and not changing position, therefore searching time may be longer for a novice user but it diminishes as user becomes familiar with interface and memorizes positions of characters. Since selection region is outside the characters region, user can dwell for much longer time without producing false selections. This major problem in interfaces with dwelling mechanism is removed. Hence a novice user can take longer searching time without making errors.

Another problem in Quickpie is that due low accuracy of the eye tracking systems and jittery movements, cursor may oscillate several times between characters and selection region which produces erroneous repetitions of selected character although user traversed correct path through gaze only one time. This problem justifies use of safe region (see figure 3). Width of safe region has to be selected according to accuracy of eye tracking. It should be less for more accurate eye tracking systems and ideally it can be zero if system works perfectly.

## EXPERIMENT

### Apparatus

The study took place using Dell-PC with 2.2 GHz Intel Core2Duo processor and 2 GB RAM under windows XP. The interface was displayed on Samsung LCD color display having resolution of 1364×768 in normal artificial lightning. Modified Sony PlayStation eye webcam (original lens was replaced by other lens without infrared (IR) filter), open source ITU gaze tracker (developed by IT University of Copenhagen) and an IR lamp consisting of a matrix of 10 IR LED were used. The webcam kept stationary at 40 cm in front of the monitor (see figure 4).

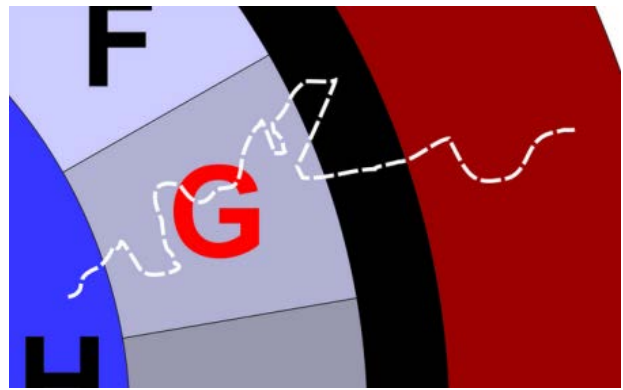

**Figure 3: Use of safe region (white curve is showing path of cursor. If safe region is not used then G will be entered two times as cursor is crossing outer edge of characters area two times.)**

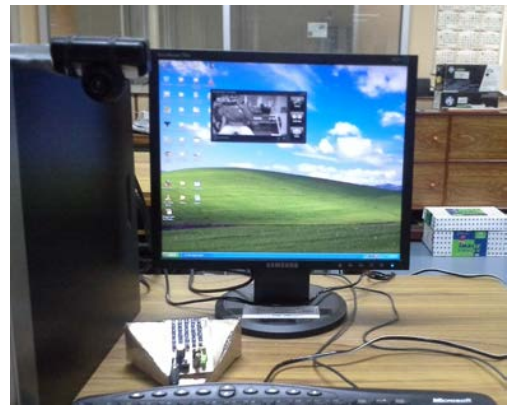

**Figure 4: Apparatus**

### Participants
Eight participants (6 male and 2 female aged between 22 and 28) participated, 2 were expert and 6 were novice to the interface. All of them except one reported normal or corrected-to-normal vision and all were familiar to computer and text composition (see figure 5).

### Procedure
Before starting experiment, user needs to calibrate eye using gaze tracker to be able to move cursor with eye gaze. Since recalibration may be required if user changes position of head, participants were asked to remember the phrase to be typed. All were explained clearly how to type using the interface. Participants were instructed to compose text as soon as possible with minimum errors and correct an error using CLEAR key before retype.

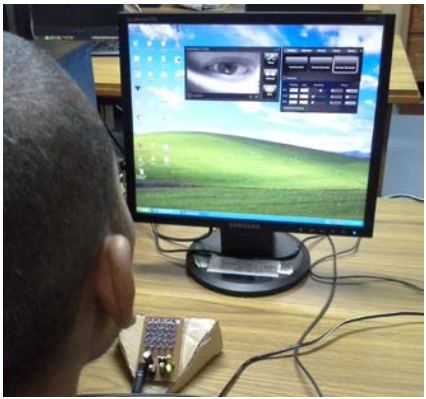

**Figure 5: Participant performing experiment**

### RESULTS
Text entry rate is measured in words per minute (wpm) and error rate in percentage of uncorrected characters to the total characters typed, taking mean of all eight users' performance.

### Number of slices:
To experiment for number of slices, designs with four, five, six and seven slices all with width of characters area 100 px were compared.

*Text entry rate*
2.53, 2.82, 3.46 and 3.08 wpm are text entry rates of designs with four, five, six and seven number of slices respectively (see figure 6).

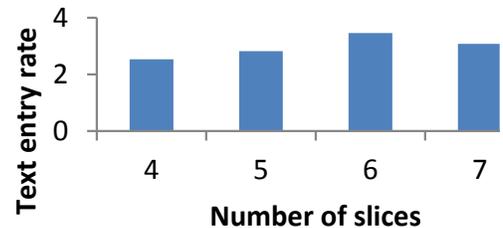

**Figure 6: Effect of number of slices on text entry rate**

*Error rate*
3.4, 2.9, 3.1 and 3.2 % are error rates of designs with four, five, six and seven number of slices respectively (see figure 7).

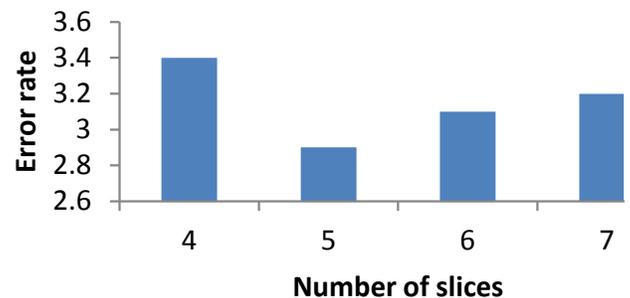

**Figure 7: Effect of number of slices on error rate**

Design with six number of slices has highest text entry rate (3.46 wpm). Although it has slightly more error rate than design with five number of slices, it is selected for further experiments.

### Width of character area:
To experiment for width of characters area, designs with width of characters area 80, 100, 120 and 140 px all with six number of slices were compared.

*Text entry rate*
3.20, 3.73, 4.27 and 3.93 wpm are text entry rates of designs with width of characters area 80, 100, 120, 140 px respectively (see figure 8).

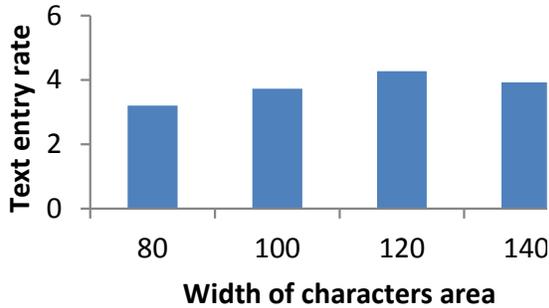

Figure 8: Effect of width of characters area on text entry rate

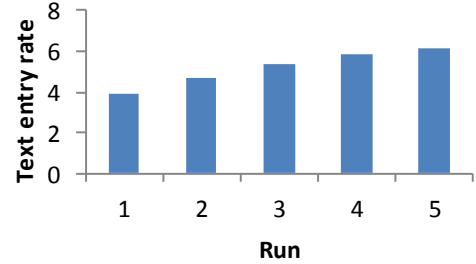

Figure 10: Effect of learnability on text entry rate

*Error rate*
3.4, 3.1, 2.7 and 2.8% are error rates of designs with width of characters area 80, 100, 120, 140 px respectively (see figure 9).

Results are showing that design with width of characters area 120 px has highest text entry rate and lowest error rate. Therefore it is best design among all other designs and selected for further experiments.

*Error rate*
2.8, 2.6, 2.7, 2.9 and 2.8 % are error rates of first, second, third fourth and fifth run respectively (see figure 11).

Text entry rate is increasing faster in early runs but it is not increasing significantly in later runs. Also there is not any pattern in improvement of error rate in subsequent runs.

**Selection method:**
The design with width of characters area 120 px and six number of slices was taken again with dwelling time selection method. Dwell time of 400 ms was taken to compare results with selection area method. Text entry rate of 4.38 wpm and error rate of 3.4% were measured. As compared to selection area method in previous experiment, dwelling time selection method is worse in both text entry rate and error rate.

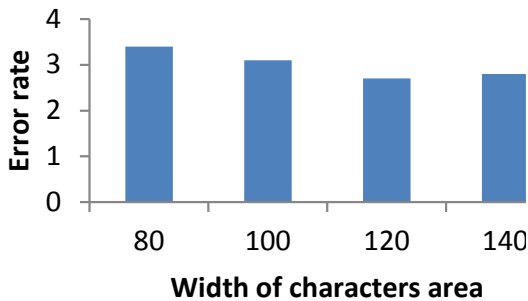

Figure 9: Effect of width of characters area on error rate

**Learnability:**
To experiment learnability, design with six number of slices and width of characters area 120 px was taken. One run with this design was held in previous experiment and again experiments with same design were repeated four times.

*Test entry rate*
3.93, 4.64, 5.39, 5.85 and 6.14 wpm are text entry rates of first, second, third fourth and fifth run respectively (see figure 10).

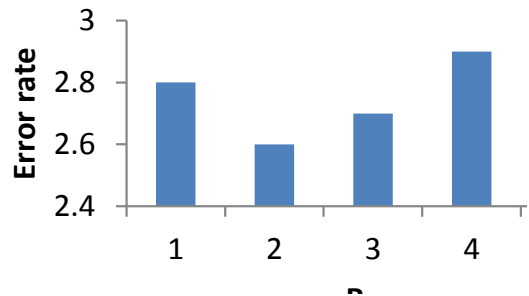

Figure 11: Effect of learnability on error rate

**DISCUSSION AND CONCLUSION**
The proposed interface is having no timing threshold as dwell time is removed. It is taking advantage of mark ahead selection that is why performance increases significantly as user gets familiar with the interface. To achieve higher text entry rate, various parameters like number of slices and width of characters region are crucial factors. Also radius of pie region, width of safe region and enlarged angle may affect performance but it is not investigated. Experiment results shows that design with sis number of slices and width of characters area 120 px is best in typing speed ond error rate.

Further research can be carried out to minimize effects of jittery movements in this interface and best values of other parameters according to apparatus and type of users.